\documentclass[10pt,twocolumn,twoside]{IEEEtran}

\usepackage[dvips]{graphicx}
\usepackage{setspace,psfrag,cite,subfigure, amsfonts,theorem}
\usepackage{amssymb}
\usepackage{bm}
\usepackage[french,english]{babel}
\usepackage{color}
\usepackage{amsmath}
\usepackage{algorithmic}
\usepackage{algorithm}

\ifCLASSINFOpdf
\else
\fi

\begin{document}
%
\title{Sparsity Averaging for Compressive Imaging}

\author{Rafael~E.~Carrillo,
        Jason D. McEwen,
        Dimitri Van De Ville,
        Jean-Philippe Thiran, 
        and Yves Wiaux

\thanks{REC is supported by the Swiss National Science Foundation (SNSF) under grant 200021-130359. JDM is supported by a Newton International Fellowship from the Royal Society and the British Academy. YW is supported by the Center for Biomedical Imaging (CIBM) of the Geneva and Lausanne Universities and EPFL, and by the SNSF under grant PP00P2-123438.}
\thanks{REC, DVDV, JPT and YW are with the Institute of Electrical Engineering, Ecole Polytechnique F{\'e}d{\'e}rale de Lausanne (EPFL),
      CH-1015 Lausanne, Switzerland. 
      DVDV and YW are also with the Department of Radiology and Medical Informatics, University of Geneva (UniGE), 
      CH-1211 Geneva, Switzerland. 
      JDM is with the Department of Physics and Astronomy, University College London, London WC1E 6BT, U.K. 
      
      E-mail: rafael.carrillo@epfl.ch (REC).}%

}%

\maketitle

\begin{abstract}
We discuss a novel sparsity prior for compressive imaging in the context of the theory of compressed sensing with coherent redundant dictionaries, based on the observation that natural images exhibit strong average sparsity over multiple coherent frames. We test our prior and the associated algorithm, based on an analysis reweighted $\ell_1$ formulation, through extensive numerical simulations on natural images for spread spectrum and random Gaussian acquisition schemes. Our results show that average sparsity outperforms state-of-the-art priors that promote sparsity in a single orthonormal basis or redundant frame, or that promote gradient sparsity. Code and test data are available at https://github.com/basp-group/sopt.
\end{abstract}


\section{Introduction}
\label{sec:intro}
Compressed sensing (CS) introduces a signal acquisition framework that goes beyond the traditional Nyquist sampling paradigm~\cite{fornasier11}. Consider a complex-valued signal $\bm{x}\in\mathbb{C}^{N}$, assumed to be sparse in some orthonormal basis $\mathsf{\Psi}\in\mathbb{C}^{N\times N}$, i.e.~$\bm{x} = \mathsf{\Psi}\bm{\alpha}$ for $\bm{\alpha}\in\mathbb{C}^{N}$ sparse. Also consider the measurement model $\bm{y}=\mathsf{\Phi}\bm{x}+\bm{n}$, where $\bm{y}\in\mathbb{C}^{M}$ denotes the measurement vector, $\mathsf{\Phi}\in\mathbb{C}^{M\times N}$ with $M<N$ is the sensing matrix, and $\bm{n}\in\mathbb{C}^{M}$ represents noise. The most common approach to recover $\bm{x}$ from $\bm{y}$ is to solve the following convex problem \cite{fornasier11}: $\min_{\bar{\bm{\alpha}}\in\mathbb{C}^{N}}\|\bar{\bm{\alpha}}\|_{1}
\textnormal{ subject to }\| \bm{y}-\mathsf{\Phi \Psi}\bar{\bm{\alpha}}\|_{2}\leq\epsilon$, where $\epsilon$ is an upper bound on the $\ell_{2}$ norm of the noise and $\|\cdot\|_1$ denotes the $\ell_{1}$ norm. 
The signal is recovered as $\hat{\bm{x}}=\mathsf{\Psi}\hat{\bm{\alpha}}$, where $\hat{\bm{\alpha}}$ denotes the solution to the above problem. Such problems, solving for the signal representation in a sparsity basis, are known as synthesis-based problems. Standard CS provides results if $\mathsf{\Phi}$ obeys a Restricted Isometry Property (RIP) and $\mathsf{\Psi}$ is orthonormal~\cite{fornasier11}. However, signals often exhibit better sparsity in a redundant dictionary ~\cite{gribonval03,bobin07,starck10}. 

Recent works have begun to address CS with redundant dictionaries, i.e.~where $\mathsf{\Psi}\in\mathbb{C}^{N\times D}$, with $N<D$, so that $\bm{x} = \mathsf{\Psi}\bm{\alpha}$ with $\bm{\alpha}\in\mathbb{C}^{D}$. Rauhut et al.~\cite{rauhut08} find conditions on $\mathsf{\Psi}$ such that $\mathsf{\Phi \Psi}$ obeys the RIP to recover $\bm{\alpha}$ in a synthesis formulation. Cand\`{e}s et al.~\cite{candes10} provide a theoretical analysis of the $\ell_1$ analysis-based problem. As opposed to synthesis, the analysis formulation solves for the signal itself:
\begin{equation}\label{cs6}
\min_{\bar{\bm{x}}\in\mathbb{C}^{N}}\|\mathsf{\Psi}^{\dagger}\bar{\bm{x}}\|_{1}
\textnormal{ subject to }\| \bm{y}-\mathsf{\Phi}\bar{\bm{x}}\|_{2}\leq\epsilon,
\end{equation}
where $\mathsf{\Psi}^{\dagger}$ denotes the adjoint operator of $\mathsf{\Psi}$. The aforementioned work~\cite{candes10} extends the standard CS theory to coherent and redundant dictionaries, providing theoretical stability guarantees based on a general condition of the sensing matrix $\mathsf{\Phi}$, coined the Dictionary Restricted Isometry Property (D-RIP). The D-RIP is a natural extension of the standard RIP. In fact many random matrices that obey the standard RIP also obey the D-RIP, like Gaussian or Bernoulli ensembles. Also, the subsampled Fourier matrix multiplied by a random sign matrix satisfies the D-RIP~\cite{krahmer11}, which provides a fast sensing operator. Interestingly, this approach falls within the spread spectrum framework proposed in \cite{puy12b}. If $\mathsf{\Phi}$ satisfies the D-RIP and $\mathsf{\Psi}$ is a general frame, Cand\`{e}s et al.\ prove in \cite{candes10} that the solution to \eqref{cs6}, denoted $\hat{\bm{x}}$, satisfies the following error bound:
\begin{equation}\label{thm1}
\| \hat{\bm{x}}-\bm{x}\|_2\leq C_0\epsilon + C_1K^{-1/2}\left \| \mathsf{\Psi}^{\dagger}\bm{x}-(\mathsf{\Psi}^{\dagger}\bm{x})_K \right\|_1,
\end{equation}
where $(\mathsf{\Psi}^{\dagger}\bm{x})_K$ denotes the best $K$-term approximation of $\mathsf{\Psi}^{\dagger}\bm{x}$ and $C_0$ and $C_1$ are numerical constants. Similar properties to the D-RIP coined $\Omega$-RIP are introduced in \cite{giryes13} in the context of the co-sparsity analysis model.

In \cite{carrillo12} some of the authors of this paper proposed a novel sparsity analysis prior in the context of Fourier imaging in radio astronomy. Our approach relies on the observation that natural images are simultaneously sparse in various frames, in particular wavelet frames, or in their gradient, so that promoting average signal sparsity over multiple frames should be a powerful prior. In the present work, the average sparsity prior is put in the generic context of compressive imaging within the theory of CS with coherent redundant dictionaries. The associated reconstruction algorithm, based on an analysis reweighted $\ell_1$ formulation, is dubbed Sparsity Averaging Reweighted Analysis (SARA). We evaluate SARA through extensive numerical simulations for spread spectrum and Gaussian acquisition schemes. Our results show that the average sparsity prior outperforms state-of-the-art priors. 

\section{Sparsity Averaging Reweighted Analysis}
\label{sec:SARA}
Natural images are often complicated and include several types of structures admitting sparse representations in different frames. For example piecewise smooth structures exhibit gradient sparsity, while extended structures are better encapsulated in wavelet frames. Therefore, in \cite{carrillo12} we observed that promoting average sparsity over multiple bases rather than a single basis is an extremely powerful prior. Here, we propose using a dictionary composed of a concatenation of $q$ frames $\mathsf{\Psi}_i$ with $1\leq i\leq q$. We focus on the particular case of concatenation of Parseval frames, creating the Parseval frame $\mathsf{\Psi}\in\mathbb{C}^{N\times D}$, with $N<D$, as:
\begin{equation}
\mathsf{\Psi}=\frac{1}{\sqrt{q}}[\mathsf{\Psi}_1, \mathsf{\Psi}_2, \ldots, \mathsf{\Psi}_q].
\end{equation}
The analysis-based framework is a suitable approach to promote average sparsity and thus we propose the following prior, proportional to the average sparsity:
\begin{equation}\label{avs}
\|\mathsf{\Psi}^{\dagger}\bar{\bm{x}}\|_{0} =\sum_{i=1}^q \|\mathsf{\Psi}_i^{\dagger}\bar{\bm{x}}\|_{0}.
\end{equation}
Note that in this setting each frame contains all the signal information. Such a prior cannot be formulated in a synthesis-based perspective. Previous works considering multiple frames, e.g. \cite{gribonval03,bobin07}, consider a component separation approach, decomposing the signal as $\bm{x}=\sum_{i=1}^q\bm{x}_i$, where each component $\bm{x}_i$ is sparse in the $i$-th frame. This is a completely different problem, where each component bears only part of the signal information, which can be addressed either in an analysis or in a synthesis framework. 

Also note on a theoretical level that a single signal cannot be arbitrarily sparse simultaneously in a set of incoherent frames \cite{elad02}. For example, a signal extremely sparse in the Dirac basis is completely spread in the Fourier basis and thus \eqref{thm1} does not provide a good error bound. As discussed by Cand\`{e}s et al.\ in \cite{candes10}, what is important is that the columns of the Gram matrix $\mathsf{\Psi}^{\dagger}\mathsf{\Psi}$ are reasonably sparse such that $\mathsf{\Psi}^{\dagger}\bm{x}$ is sparse when $\bm{x}$ admits a sparse representation $\bm{\alpha}$ with $\bm{x}= \mathsf{\Psi}\bm{\alpha}$. This requirement is nothing else than a coherence condition on $\mathsf{\Psi}$. In our case of concatenations of frames, this leads to the condition that each $\mathsf{\Psi}_i$ is highly coherent with itself and mutually coherent with the other frames.  The component separation approaches in \cite{gribonval03,bobin07} use incoherent frames for the decomposition, while our average sparsity prior takes the opposite direction.
The concatenation of the first eight orthonormal Daubechies wavelet bases (Db1-Db8, $q=8$) represents a good and simple candidate for a dictionary in imaging applications. The first Daubechies wavelet basis, Db1, is the Haar wavelet basis, which can be used as an alternative to gradient sparsity (usually imposed by a total variation (TV) prior~\cite{chambolle04}) to promote piecewise smooth signals. The Db2-Db8 bases provide smoother sparse decompositions. All Daubechies bases are mutually coherent thanks to their compact support and identical sampling positions. 

In order to promote average sparsity through the prior \eqref{avs} we adopt a reweighted $\ell_1$ minimization scheme \cite{candes08a}. The algorithm replaces the $\ell_0$ norm by a weighted $\ell_1$ norm and solves a sequence of weighted $\ell_1$ problems with weights essentially the inverse of the values of the solution of the previous problem: 
\begin{equation}\label{delta}
\min_{\bar{\bm{x}}\in\mathbb{C}^{N}}\|\mathsf{W\Psi}^{\dagger}\bar{\bm{x}}\|_{1}
\textnormal{ subject to }\| \bm{y}-\mathsf{\Phi}\bar{\bm{x}}\|_{2}\leq\epsilon,
\end{equation}
where $\mathsf{W}\in\mathbb{R}^{D\times D}$ is a diagonal matrix with positive weights. Assuming i.i.d.~complex Gaussian noise with variance $\sigma_n$, the $\ell_{2}$ norm term in \eqref{delta} is identical to a bound on the $\chi^{2}$ with $2M$ degrees of freedom governing the noise level estimator. Therefore, we set this bound as
 $\epsilon^2=(2M+4\sqrt{M})\sigma_n^2/2$, where $\sigma^2_n/2$ is the variance of both the real and imaginary parts of the noise. This choice provides a likely bound for $\|\bm{n}\|_2$ \cite{carrillo12}. To solve \eqref{delta}, we use the Douglas-Rachford splitting algorithm \cite{combettes07}. The solution is denoted as $\Delta(\bm{y}, \mathsf{\Phi},\mathsf{W},\epsilon)$.
The weights are updated at each iteration, i.e. after solving a complete weighted $\ell_1$ problem, by the function $f(\gamma,a)=\gamma(\gamma+|a|)^{-1}\in (0,1]$, where $a$ denotes the coefficient value estimated at the previous iteration and $\gamma\neq 0$ plays the role of a stabilization parameter, avoiding undefined weights when the signal value is zero. Note that as $\gamma\rightarrow 0$ the solution of the weighted $\ell_1$ problem approaches the solution of the $\ell_0$ problem. We use a homotopy strategy and solve a sequence of weighted $\ell_1$ problems using a decreasing sequence $\{\gamma^{(t)}\}$, with $t$ denoting the iteration time variable. The resulting algorithm, dubbed sparsity averaging reweighted analysis (SARA), is defined in Algorithm~\ref{alg1}\footnote{A rate parameter $\beta \in (0,1)$ controls the decrease of the sequence $\gamma^{(t)}=\beta\gamma^{(t-1)}$. In practice $\gamma^{(t)}$ should however not reach zero. The noise standard deviation in the sparsity domain $\sigma_{\alpha}=\sqrt{M/D}\sigma_n$, with $\sigma_n$ the noise standard deviation in measurement space, is a rough estimate for a baseline above which significant signal components could be identified. Hence we set $\gamma^{(t)}=\max\{\beta\gamma^{(t-1)},\sigma_{\alpha}\}$ so that $\gamma^{(t)}$ is lower-bounded by $\sigma_{\alpha}$. As a starting point we set $\hat{\bm{x}}^{(0)}$ as the solution of the $\ell_1$ problem and $\gamma^{(0)}=\sigma_s\left(\mathsf{\Psi}^{\dagger}\hat{\bm{x}}^{(0)}\right)$, where $\sigma_s(\cdot)$ takes the empirical standard deviation of a signal. The re-weighting process stops  when the relative variation between successive solutions is smaller than some bound $\eta\in(0,1)$, or after a maximum number of iterations $N_{\rm{max}}$. We fix $\eta=10^{-3}$ and $\beta=10^{-1}$.}. See \cite{carrillo12} for more details.

\begin{algorithm}[h!]
\caption{SARA algorithm}
\label{alg1}
\begin{algorithmic}[1]
\REQUIRE $\bm{y}$, $\mathsf{\Phi}$, $\epsilon$, $\sigma_{\alpha}$, $\beta$, $\eta$ and $N_{\rm{max}}$.
\ENSURE Reconstructed image $\hat{\bm{x}}$.
\STATE Initialize $t=1$, $\mathsf{W}^{(0)}=\mathsf{I}$ and $\rho=1$.
\STATE Compute\\
$\hat{\bm{x}}^{(0)}=\Delta(\bm{y}, \mathsf{\Phi},\mathsf{W}^{(0)},\epsilon)$,
$\gamma^{(0)}=\sigma_s\left(\mathsf{\Psi}^{\dagger}\hat{\bm{x}}^{(0)}\right)$.
\WHILE{$\rho>\eta$ and $t<N_{\rm{max}}$}
\STATE Update 
$\mathsf{W}_{ij}^{(t)}=f\left (\gamma^{(t-1)},\hat{\alpha}_{i}^{(t-1)}\right)\delta_{ij}$, \\
for $i,j=1,\ldots,D$ with $\hat{\bm{\alpha}}^{(t-1)}=\mathsf{\Psi}^{\dagger}\hat{\bm{x}}^{(t-1)}$.
\STATE Compute a solution
$\hat{\bm{x}}^{(t)}=\Delta(\bm{y}, \mathsf{\Phi},\mathsf{W}^{(t)},\epsilon)$.\\
\STATE Update $\gamma^{(t)}=\max\{\beta\gamma^{(t-1)},\sigma_{\alpha}\}$.
\STATE Update $\rho=\| \hat{\bm{x}}^{(t)}-\hat{\bm{x}}^{(t-1)}\|_2/\|\hat{\bm{x}}^{(t-1)}\|_2$
\STATE $t\leftarrow t+1$
\ENDWHILE
\end{algorithmic}
\end{algorithm}

\section{Experimental Results}
\label{sec:Simulations and results}

\begin{figure*}
\centering
   
    \subfigure[]{\includegraphics[trim = 0.8cm 0.1cm 1.5cm 0.8cm, clip, keepaspectratio, width = 4.4cm]{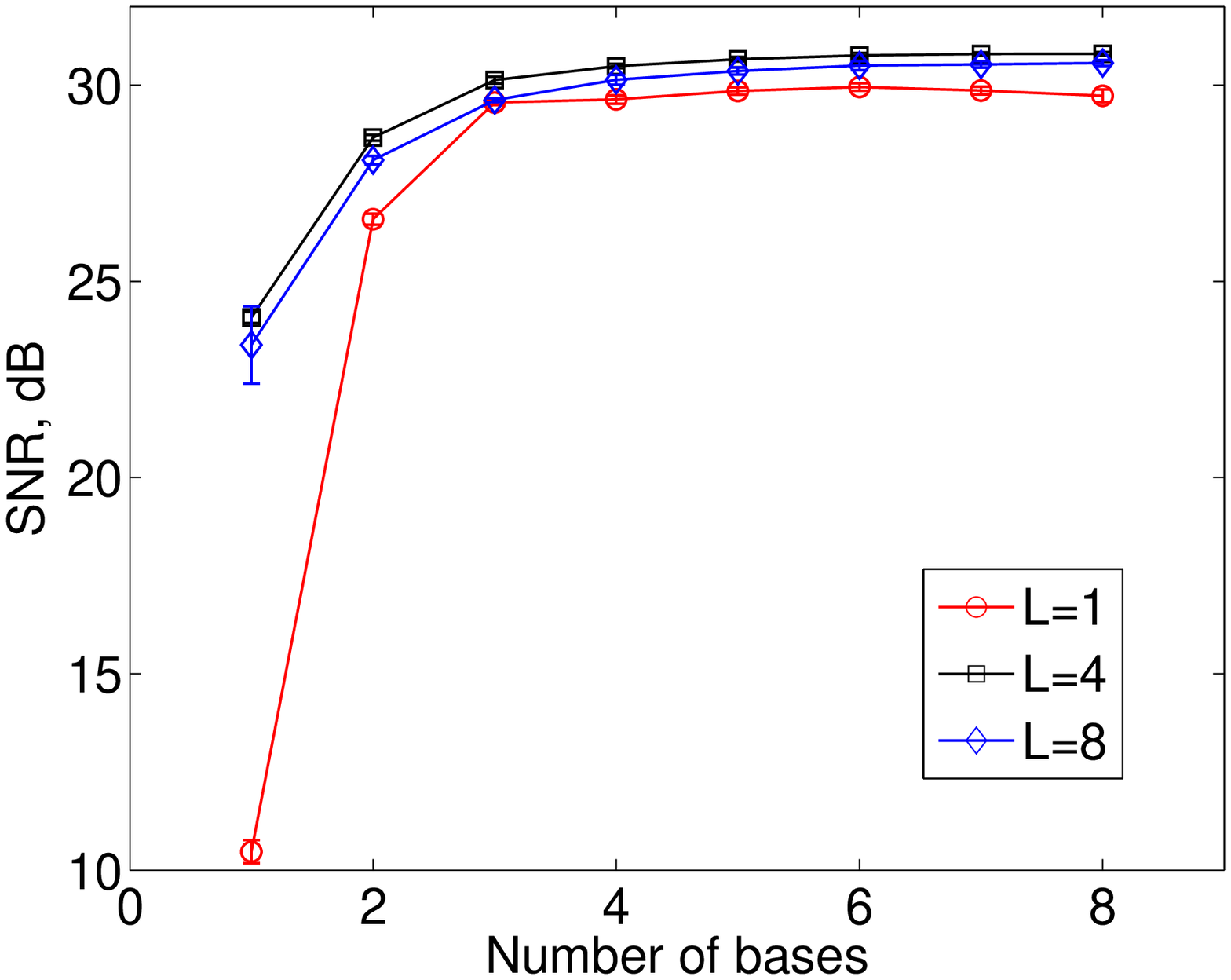}\label{fig:1a}}
    \subfigure[]{\includegraphics[trim = 0.8cm 0.1cm 1.5cm 0.8cm, clip, keepaspectratio, width = 4.4cm]{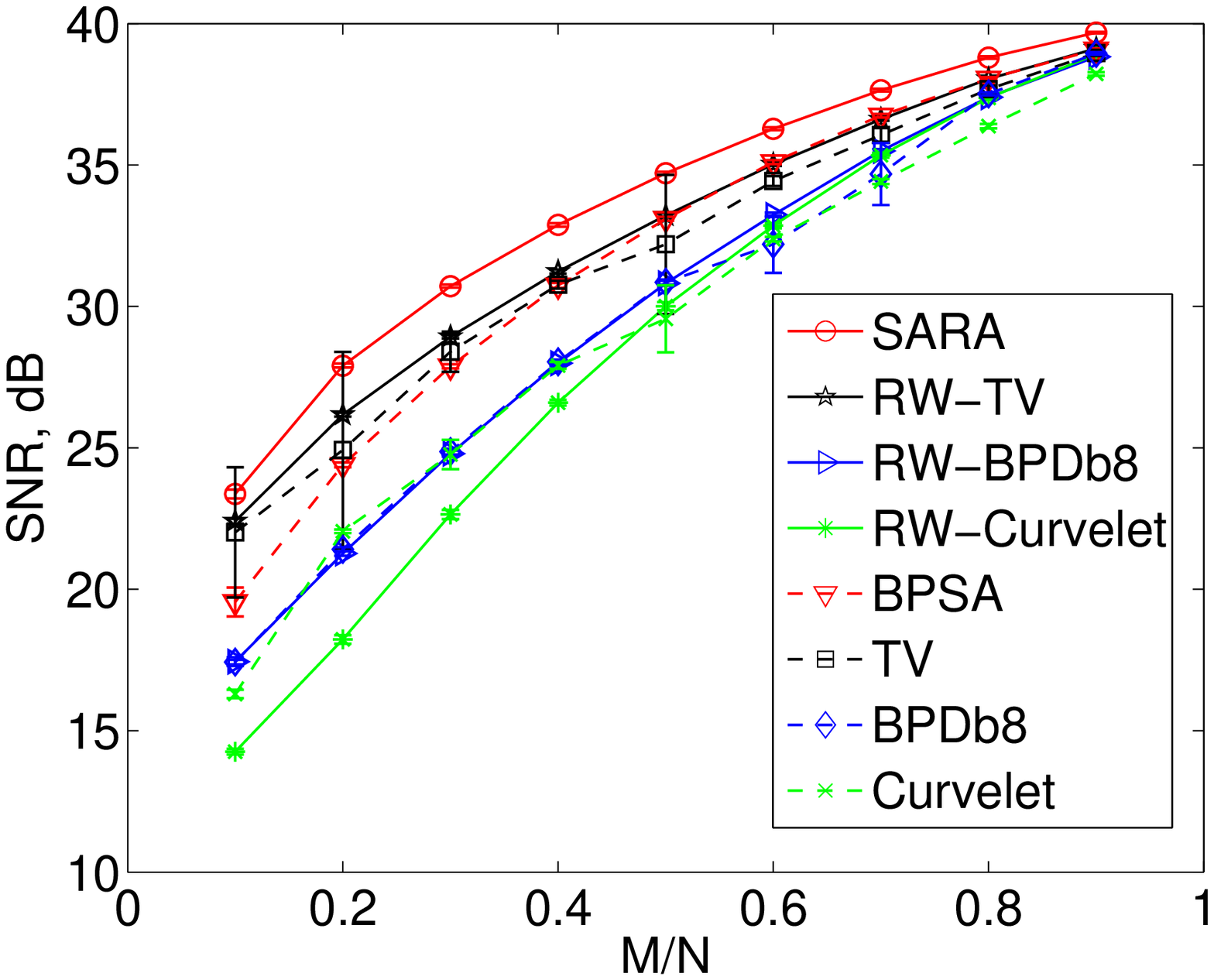}\label{fig:1b}}
    \subfigure[]{\includegraphics[trim = 0.8cm 0.1cm 1.5cm 0.8cm, clip, keepaspectratio, width = 4.4cm]{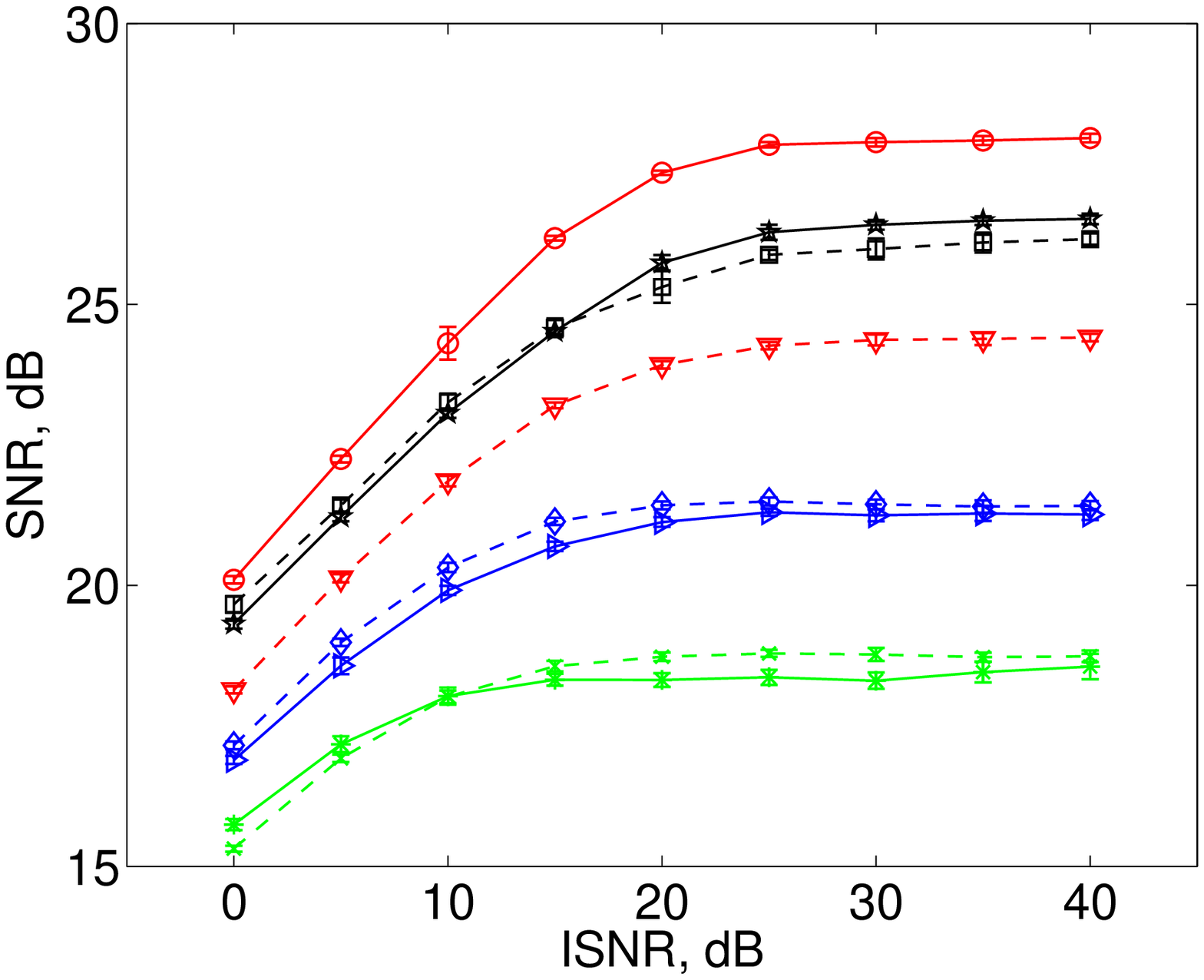}\label{fig:1c}}
    \subfigure[]{\includegraphics[trim = 0.8cm 0.1cm 1.5cm 0.8cm, clip, keepaspectratio, width = 4.4cm]{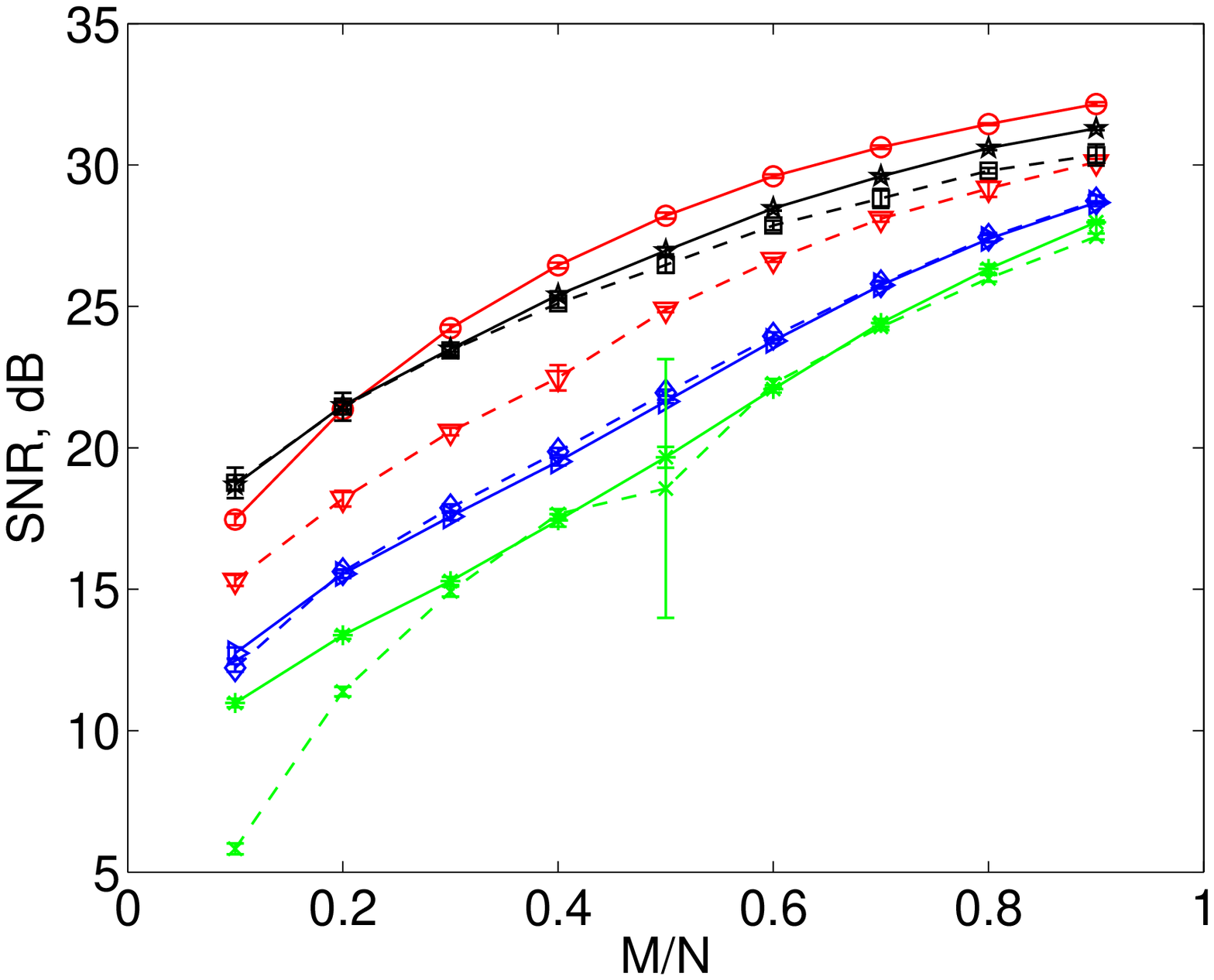}\label{fig:1d}}
  
\caption{Reconstruction quality results for Lena and spread spectrum measurements. (a) $\mathrm{SNR}$ as a function of the number of bases in the dictionary for decomposition depths $L=1,4,8$ ($M=0.3N$, $\mathrm{ISNR}=30$~dB). (b) $\mathrm{SNR}$ results against the undersampling ratio ($\mathrm{ISNR}=30$~dB).  (c) $\mathrm{SNR}$ as a function of $\mathrm{ISNR}$ ($M=0.2N$).  (d) Results for random random Gaussian measurements. $\mathrm{SNR}$ against the undersampling ratio for cropped Lena image ($\mathrm{ISNR}=30$~dB).}
\label{fig:1}
\end{figure*}

In this section we evaluate the reconstruction performance of SARA by recovering a 256$\times$256 pixel version of the Lena test image from compressive measurements following the measurement model presented in Section \ref{sec:intro}. We use the suggested Db1-Db8 concatenation as the dictionary for SARA. In order to have a fast measurement operator that obeys the D-RIP, we use for a first experiment the spread spectrum technique described in \cite{puy12b}. Spread spectrum incorporates a modulating sequence on top of Fourier sampling, defining the measurement operator as $\mathsf{\Phi}=\mathsf{MF}\mathsf{C}$, where $\mathsf{C}\in\mathbb{R}^{N\times N}$ is a diagonal matrix with elements with unit norm and randomized sign, $\mathsf{F}\in\mathbb{C}^{N\times N}$ is the discrete Fourier operator and $\mathsf{M}\in\mathbb{R}^{M\times N}$ is a binary mask defining the random selection operator.  For a second experiment we consider Gaussian random measurement matrices.

We compare SARA to analogous analysis algorithms, and their reweighted versions, changing the sparsity dictionary $\mathsf{\Psi}$ in \eqref{cs6} and \eqref{delta} respectively. The three different dictionaries are: the Daubechies 8 wavelet basis, the redundant curvelet frame~\cite{starck10} and the Db1-Db8 concatenation. The associated algorithms are respectively denoted BPDb8, Curvelet and BPSA for the non reweighted case. The reweighted versions are respectively denoted RW-BPDb8, RW-Curvelet and SARA. We also compare to the TV prior~\cite{chambolle04}, where the TV minimization problem is formulated as a constrained problem like \eqref{cs6}, but replacing the $\ell_1$ norm by the image TV norm. The reweighted version of TV is denoted as RW-TV. Since the image of interest is positive, we impose the additional constraint that $\bar{\bm{x}}\in\mathbb{R}_{+}^N$ for all problems.

We use as reconstruction quality metric the standard signal-to-noise ratio ($\mathrm{SNR}$), defined as $\mathrm{SNR}=20\log_{10}\left( \|\bm{x}\|_2/\|\bm{x}-\hat{\bm{x}}\|_2\right)$, where $\bm{x}$ and $\hat{\bm{x}}$ denote the original and the estimated image respectively. Average values over 30 simulations and associated $1\sigma$ error bars are reported for all experiments. The measurements are corrupted by complex Gaussian noise. The associated input $\mathrm{SNR}$ is defined as $\mathrm{ISNR}=20\log_{10}(\| \bm{y}_0\|_2/\| \bm{n}\|_2)$, where $\bm{y}_0$ identifies the clean measurement vector.

We start by evaluating SARA for spread spectrum acquisition. Prior to our main analysis, we study the reconstruction performance of SARA as a function of the number of wavelet bases in the dictionary. We test depths $L=1,4,8$ in the Daubechies decomposition for all dictionaries, fixing $M=0.3N$ and $\mathrm{ISNR}=30$~dB. We add bases in parametric order, i.e., one basis means Db1 alone, two bases Db1 and Db2 and so on until we reach the eight bases from Db1-Db8. The results for Lena are summarized in Figure~\ref{fig:1a}. We can observe that the best performance is obtained when $L=4$ and the worst when $L=1$. We can also observe that the reconstruction quality improves as the number of bases increases until it saturates between 4 to 8 bases. These results corroborate our choice for 8 bases, and $L=4$.

Having validated the dictionary choice, we now proceed to evaluate the reconstruction quality of SARA as a function of the undersampling ratio $M/N$. We fix $\mathrm{ISNR}=30$~dB and vary the undersampling ratio from 0.1 to 0.9. The $\mathrm{SNR}$ results comparing SARA against all the other benchmark methods are shown in Figure \ref{fig:1b}. The results demonstrate that SARA outperforms state-of-the-art methods for all undersamplings. SARA achieves gains between 0.9 and 1.9~dB with the largest gains observed for undersampling ratios in the range 0.2-0.5. Notably, BPSA achieves better $\mathrm{SNR}$ than BPDb8, curvelet and their reweighted versions for all undersampling ratios. It also achieves similar $\mathrm{SNR}$ to TV in the range 0.4-0.9.

The following experiment studies the robustness of SARA against measurement noise in the spread spectrum acquisition setting. We fix $M=0.2N$ and vary the $\mathrm{ISNR}$ in the range 0 to 40~dB. The results are summarized in Figure \ref{fig:1c}. As expected from the bound in \eqref{thm1}, the relationship between $\mathrm{SNR}$ and  $\mathrm{ISNR}$ is linear with slope 1 for low $\mathrm{ISNR}$ until it is high enough and the reconstruction quality is dominated by  the undersampling effect. Notably, SARA outperforms the benchmark methods for all $\mathrm{ISNR}$, achieving an $\mathrm{SNR}$ of 20~dB for an $\mathrm{ISNR}$ of 0~dB. Again, BPSA yields a better performance than BPDb8, Curvelet and their reweighted versions.

\begin{figure}[h]

\centering
    
     \includegraphics[trim = 4.3cm 1.5cm 3.6cm 0.8cm, clip, keepaspectratio, width = 4.1cm]{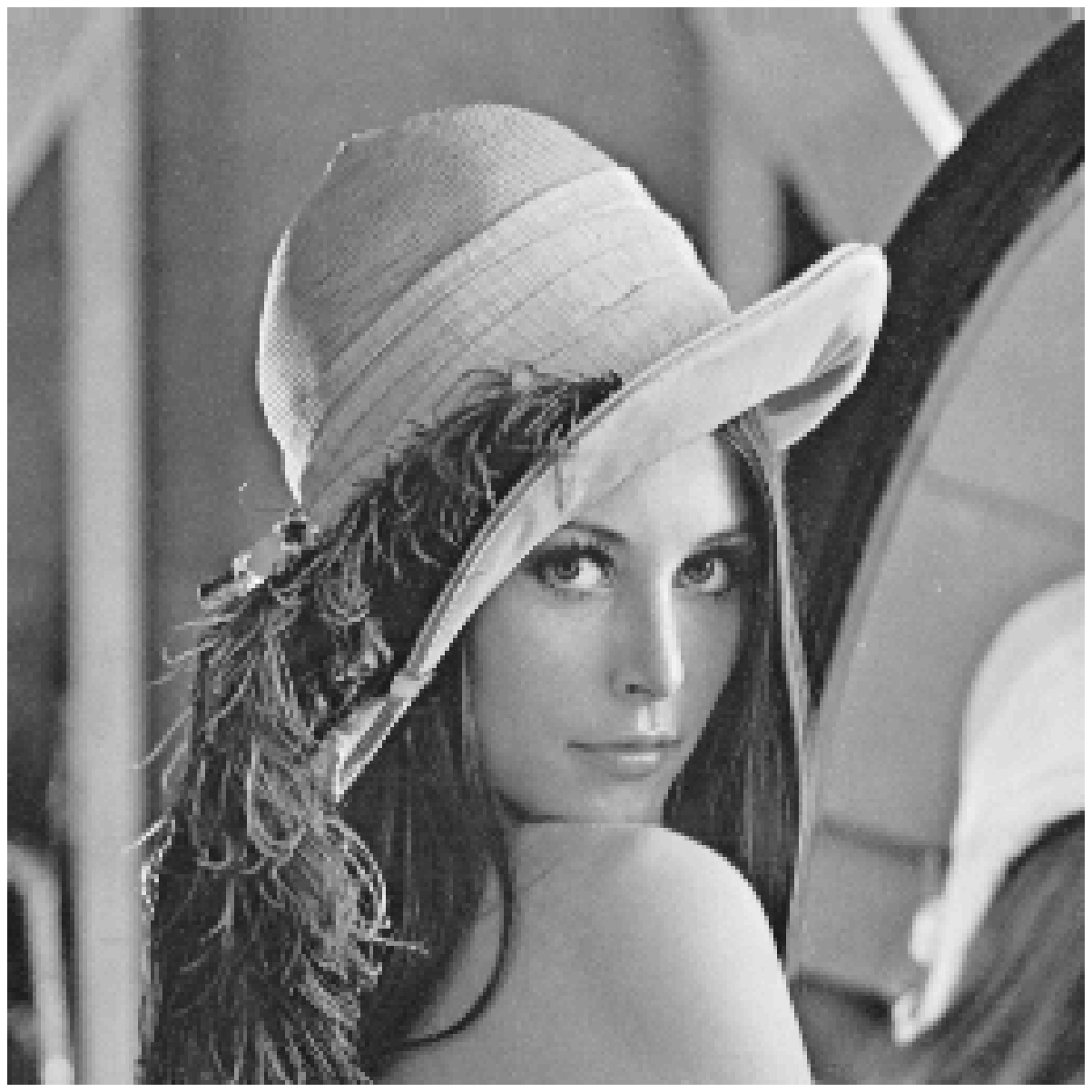}
    \includegraphics[trim = 4.3cm 1.5cm 3.6cm 0.8cm, clip, keepaspectratio,  width = 4.1cm]{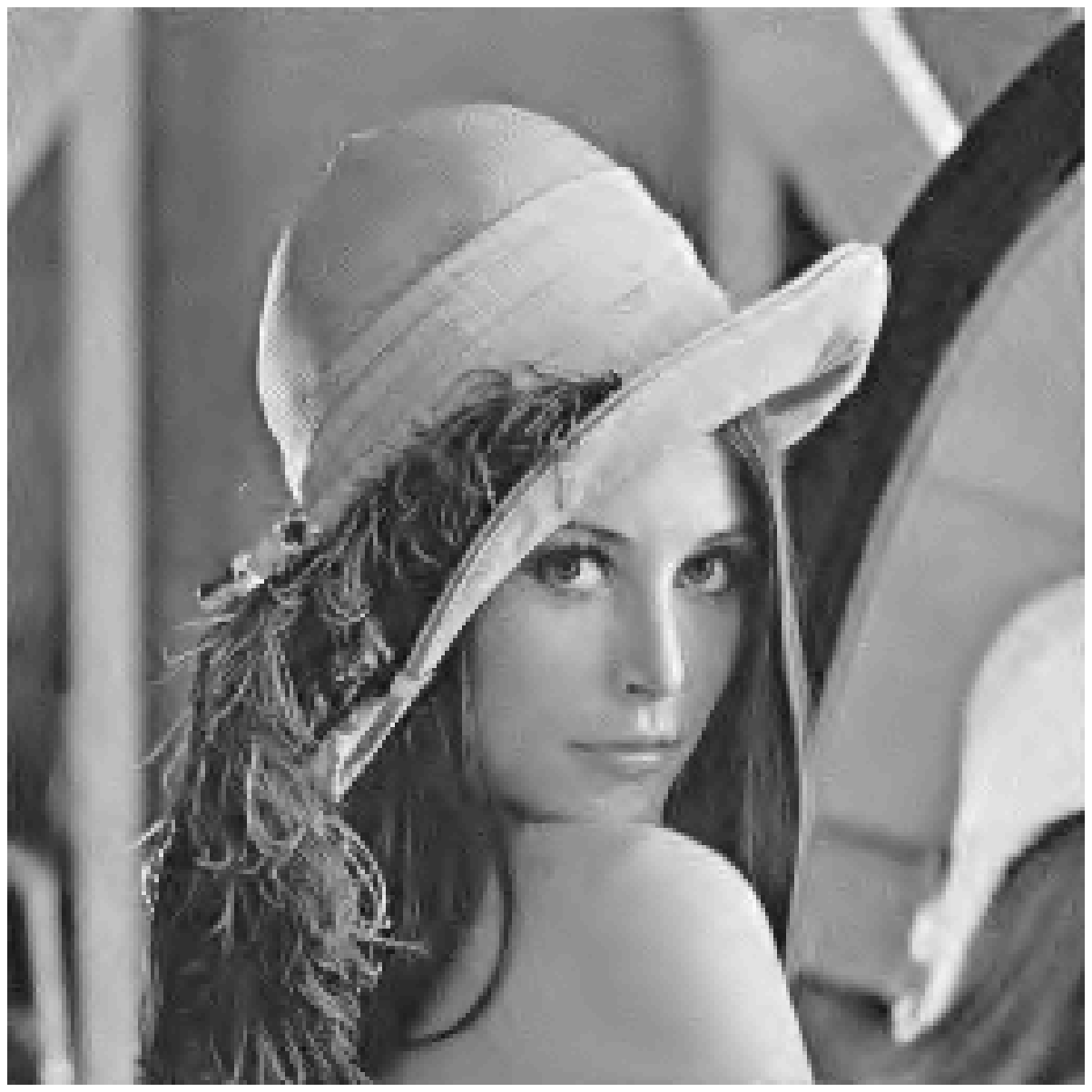}
    \includegraphics[trim = 4.3cm 1.5cm 3.6cm 0.8cm, clip, keepaspectratio,  width = 4.1cm]{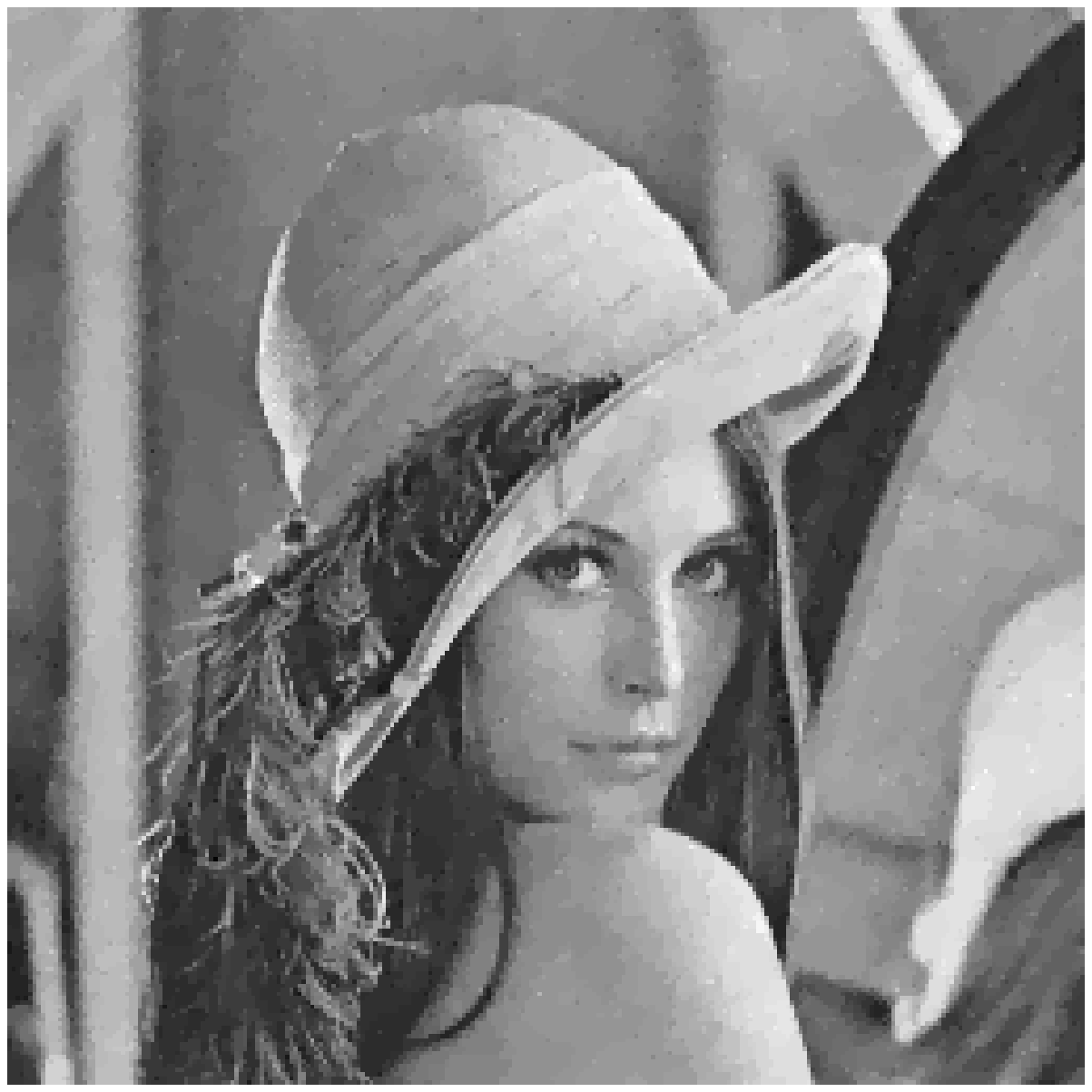}
    \includegraphics[trim = 4.3cm 1.5cm 3.6cm 0.8cm, clip, keepaspectratio,  width = 4.1cm]{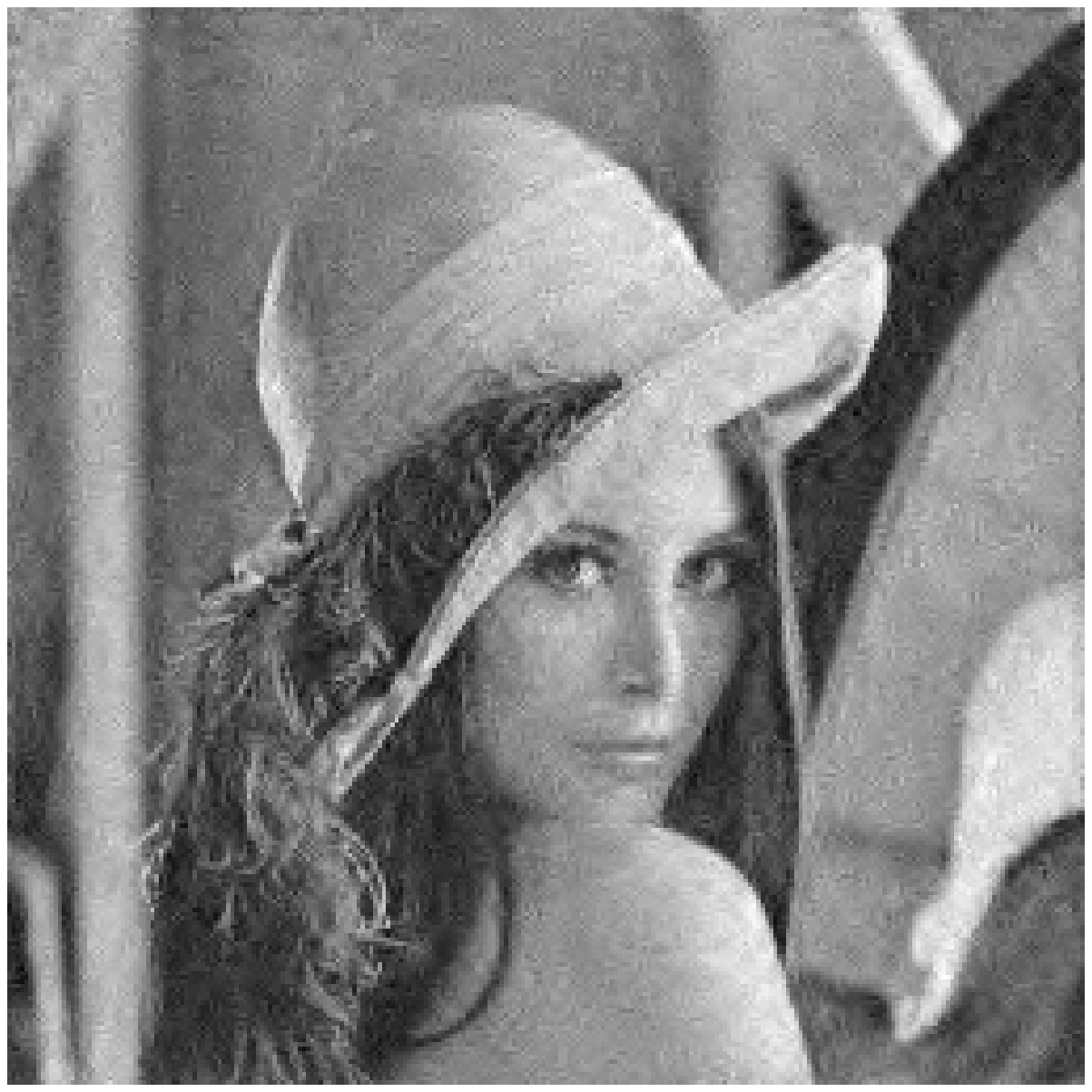}
    
\caption{Reconstruction example for Lena in spread spectrum acquisition setting ($M=0.2N$, $\mathrm{ISNR}=30$~dB). From left to right and top to bottom: original image, reconstructed images for SARA (28.1~dB), RW-TV (26.3~dB) and BPDb8 (21.4~dB).}%
\label{fig:2}
\end{figure}

Next we present a visual assessment of the reconstruction quality of SARA compared to the benchmark methods, still in the spread spectrum acquisition setting. Figure~\ref{fig:2} shows the reconstructions for $M=0.2N$ and $\mathrm{ISNR}=30$~dB for the three best algorithms in $\mathrm{SNR}$: SARA (28.1~dB), RW-TV (26.3~dB) and BPDb8 (21.4~dB). SARA provides an impressive reduction of visual artifacts relative to the other methods in this high undersampling regime. In particular RW-TV exhibits expected cartoon-like artifacts. BPDb8 does not yield results of comparable visual quality. 

We now study the performance of SARA with Gaussian random matrices as measurements operators. Due to computational limitations for the use of a dense sensing matrix, for this experiment we use a cropped version of Lena, around the head, of dimension 128$\times$128 as test image. We compare SARA against all the benchmark methods for this sensing modality. We fix $\mathrm{ISNR}=30$~dB and vary the undersampling ratio in the range 0.1 to 0.9. The $\mathrm{SNR}$ results are reported in Figure \ref{fig:1d}. These results confirm the performance of SARA for compressive imaging with a different sensing matrix, outperforming the benchmark methods for $M\geq0.3N$. For $M=0.1N$ SARA is 1~dB below TV and RW-TV and for $M=0.2N$ it achieves the same $\mathrm{SNR}$. 

\begin{figure}[h]

\centering
    
    \includegraphics[trim = 4.8cm 1.6cm 4.1cm 0.8cm, clip, keepaspectratio, width = 2.7cm]{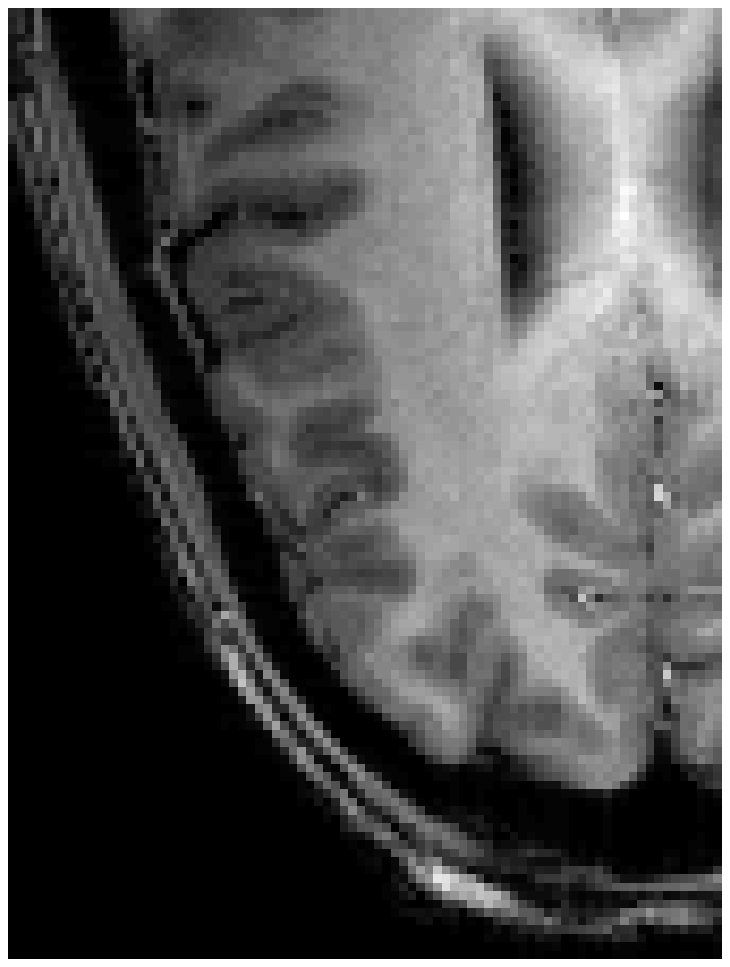} 
    \includegraphics[trim = 4.8cm 1.6cm 4.1cm 0.8cm, clip, keepaspectratio, width = 2.7cm]{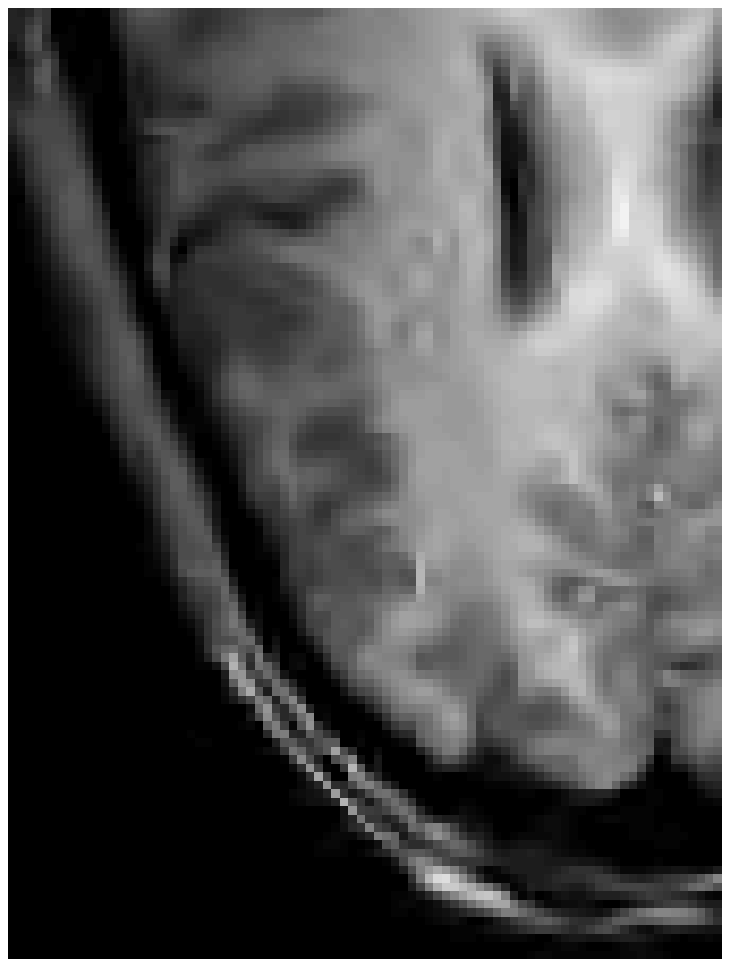}
    \includegraphics[trim = 4.8cm 1.6cm 4.1cm 0.8cm, clip, keepaspectratio, width = 2.7cm]{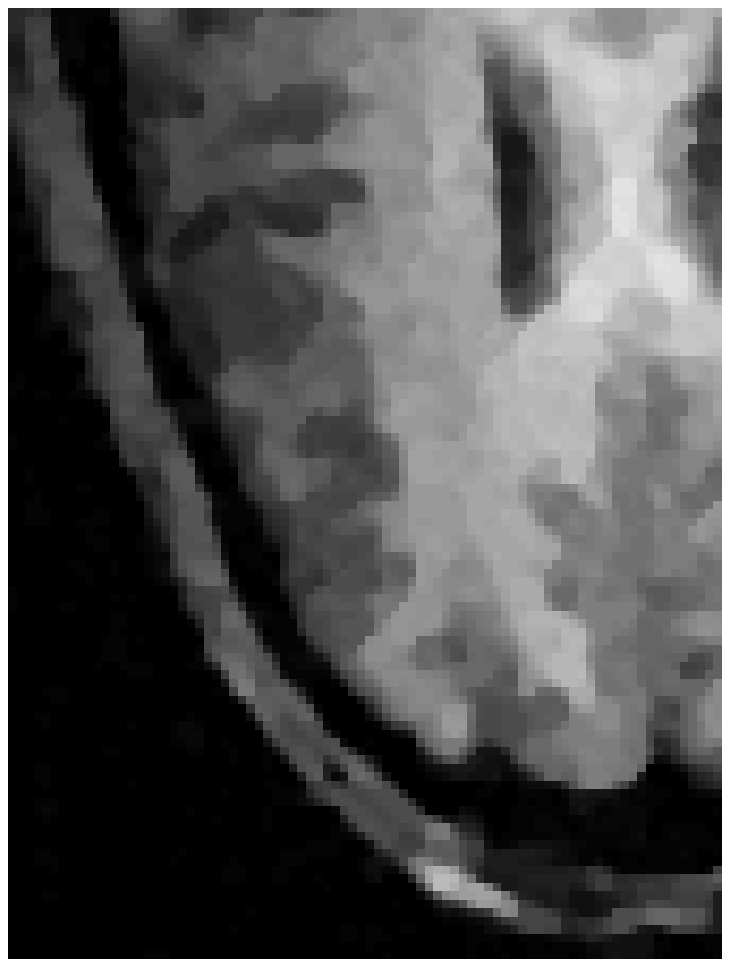}
   
\caption{MR illustration: reconstruction of a brain image from Fourier acquisition ($M=0.05N$, $\mathrm{ISNR}=30$~dB). From left to right: original image, SARA (18.8~dB) and TV (17.3~dB) reconstructions.}
\label{fig:3}
\end{figure}

As final experiment, we present a magnetic resonance (MR) imaging illustration. We reconstruct a 224$\times$168 positive brain image from standard variable density Fourier measurements, for an adverse undersampling ratio of $M=0.05N$, well beyond current state of the art in the field. The $\mathrm{ISNR}$ is set to 30~dB. In this case, the sparsity dictionary for SARA is augmented with the Dirac basis as the brain is quite localized in the field of view. Figure~\ref{fig:3} shows a zoom of the original brain image and reconstructed images for SARA and TV, which yield the two best reconstructions in $\mathrm{SNR}$. In addition to an $\mathrm{SNR}$ gain of 1.5~dB, SARA achieves an impressively better reconstruction from the visual standpoint.

\section{Conclusion}
\label{sec:Conclusion}
In this letter we have discussed the novel SARA regularization method and algorithm for compressive imaging in the theoretical context of CS with coherent redundant dictionaries. The approach relies on the observation that natural images exhibit strong average sparsity. We have evaluated SARA under two different acquisition schemes: spread spectrum and random Gaussian measurements. Experimental results demonstrate that the sparsity averaging prior embedded in the analysis reweighted $\ell_1$ formulation of SARA outperforms state-of-the-art priors, based on single frame or gradient sparsity, both in terms of $\mathrm{SNR}$ and visual quality. An MR imaging illustration also corroborates these conclusions for Fourier imaging. Code and test data are available at https://github.com/basp-group/sopt.

Future work will concentrate on finding a theoretical framework for the average sparsity model. Specialized results are indeed needed in the particular case of concatenation of frames for an estimate of the number of measurements required for accurate image reconstruction. It would be interesting to explore the connections between average sparsity and the co-sparsity model, which proposes a general framework for general analysis operators (see \cite{giryes13} and references therein).
Also, it was recently shown in \cite{oymak13} that combinations of convex relaxation priors do not yield better results than exploiting only one of those priors, while non-convex approaches can exploit multiple models. Those results suggest that the re-weighting approach in SARA to approximate the non-convex $\ell_0$ norm is fundamental to exploit average sparsity, as observed in the simulation results.

\bibliographystyle{IEEEtran}
\bibliography{abrev,sara}

\end{document}